\documentclass[seceq]{ptptex}
\usepackage{graphicx,color,amsmath}

\usepackage{geometry}
\newcommand{\Slash}[1]{\ooalign{\hfil/\hfil\crcr$#1$}}
\newcommand{\pa}{\partial}
\newcommand{\nn}{\nonumber}
\newcommand{\psibar}{{\bar \psi}}

\newcommand{\bref}[1]{(\ref{#1})}

\title{
Anomalies in the ERG approach%
}

\author{
Yuji \textsc{Igarashi},${}^1$
Katsumi \textsc{Itoh},${}^1$
Masanao \textsc{Sato}${}^2$
and Hidenori \textsc{Sonoda}${}^3$
}

\inst{
${}^1$Department of Education, Niigata University, Niigata 950-2181, Japan\\
${}^2$Graduate School of Science and Technology, Niigata University, Niigata 950-2181, Japan\\
${}^3$Physics Department, Kobe University, Kobe 657-8501, Japan
}

\abst{
The antifield formalism adapted in the exact renormalization group is found
to be useful for describing a system with some symmetry, especially the
gauge symmetry.  In the formalism, the vanishing of the quantum master
operator implies the presence of a symmetry.  The QM operator satisfies
a simple algebraic relation that will be shown to be related to the
Wess-Zumino condition for anomalies.  We also explain how an anomaly
contributes to the QM operator.  }

\begin{document}

\maketitle

\section{Introduction}

In the exact renormalization group, it often happens that the symmetry
of a system is not compatible with the momentum cutoff.  This is
particularly important for a gauge theory since we do not have a
convenient way of regularizing the theory without breaking the gauge
symmetry.\footnote{A regularization procedure to respect the gauge
symmetry has been proposed.  See Ref. \citen{Arnone:2006ie} and
references therein.}

As shown in earlier works,\cite{Becchi:1996an,Igarashi:1999rm} any
symmetry survives even after introducing the momentum cutoff $\Lambda$.
As the cutoff changes, the Wilson action and the symmetry transformation
change in their appearance.  Still, we may write the Ward-Takahashi (WT)
identity ${\Sigma}_{\Lambda}=0$, which may be elevated to the quantum
master equation (QME) ${\bar \Sigma}_{\Lambda}=0$ of the
Batalin-Vilkovisky (BV) antifield formalism.\cite{Batalin:1981jr}.  The
QME implies the presence of the symmetry in the system.  The QME ${\bar
\Sigma}_{\Lambda}=0$ in the limit of $\Lambda \rightarrow 0$ is found to
be equivalent to the Zinn-Justin equation.\footnote{See \S 10 and
Appendix D of Ref. \citen{review-PTPS}.} Since both equations are
manifestations of the presence of a symmetry, this correspondence is
quite natural.

With antifields, we introduce a canonical structure that has been fully
utilized in its application to the ERG.  In later sections, we find an
algebraic relation for ${\bar \Sigma}_{\Lambda}$ derived from its
definition and the canonical structure.  Since this is an algebraic
relation, it holds even if ${\bar \Sigma}_{\Lambda} \ne 0$, i.e., even
in the absence of the corresponding symmetry.
Naturally, we expect that the effective action also satisfies some
algebraic condition similar to the one for the QM operator.  Actually,
we already know such a condition, that is, the Wess-Zumino (WZ)
condition.  Therefore, in the case of ${\bar \Sigma}_{\Lambda} \ne 0$,
the QM operator must be related to an anomaly.  It is the subject of the
present paper to explain how the QM operator is related to an
anomaly.\footnote{Although we consider gauge anomalies in this paper,
our discussion may be extended to global anomalies that also have been
studied in the BV formalism\cite{Amorim:1996xp,Amorim:1998mm}.}

We will see that the QM operator is a composite operator, essentially an
anomaly, that flows under the change of the cutoff scale.  As a
composite operator, it changes the expression.  In the limit of $\Lambda
\rightarrow 0$, it is related to the well-known anomaly written for the
effective action.  The expression of the composite operator also
simplifies in the other limit of $\Lambda \rightarrow \infty$: the QM
operator becomes an anomaly times a ghost factor.  This will be shown
explicitly for an abelian theory.

The paper is organized as follows.  In the next section, we describe
some results reported earlier,\cite{review-PTPS} which are needed for
later discussion.  In this paper, we follow the
notations used in the review article.  Via the relation to the 1PI
counter object of the QM operator, we find how the QM operator tends to
the anomaly in the $\Lambda \rightarrow 0$ limit.  In \S 3 and \S 4, we
study the other limit of $\Lambda \rightarrow \infty$ for an abelian
gauge theory coupled to massless fermions.  We will see that the same result is
obtained by two different methods. The last section is devoted to
summary and discussion.  There, we point out the relation of the
Wess-Zumino condition to the algebraic condition on the QM operator.
The proof of the relation is given in the Appendix.

\section{Antifield formalism and its application to ERG}

Here, we describe some results that will be useful to understand later
discussion.

Given a classical gauge fixed action $S_{cl}[\phi]$ for a generic gauge
theory, we may write an extended action
\begin{eqnarray}
{\bar S}_{cl}[\phi,\phi^*] \equiv S_{cl}[\phi]+ \phi_A^* \delta \phi^A~.
\label{extended action}
\end{eqnarray}
The field $\phi^A$ represents the gauge, ghost, antighost,
auxiliary fields as well as possible matter fields.  The BRST
transformation is denoted as $\delta \phi^A$.  $\phi_A^*$
represents the corresponding antifield with the opposite Grassmann parity
to that of $\phi^A$.

In the space of $\phi^A$ and $\phi_A^*$, we define the canonical
structure via an antibracket: for any field variables $X$ and $Y$, we
define
\begin{eqnarray}
(X,Y) \equiv \frac{\partial^r X}{\partial \phi^A}\frac{\partial^l
 Y}{\partial \phi_A^*} -
\frac{\partial^r X}{\partial \phi_A^*}\frac{\partial^l Y}{\partial
\phi^A}~.
\label{antibracket}
\end{eqnarray}

Following the definitions \bref{extended action} and \bref{antibracket},
we obtain
\begin{eqnarray}
({\bar S}_{cl}, {\bar S}_{cl}) = 2 (\delta S_{cl} + \phi_A^* \delta^2 \phi^A)~.
\label{CME1}
\end{eqnarray}
The r.h.s. of \bref{CME1} vanishes if the action is BRST invariant and
the transformation is nilpotent.  Namely, under these two conditions,
the action ${\bar S}_{cl}$ satisfies the classical master equation
(CME): $({\bar S}_{cl}, {\bar S}_{cl}) = 0$.

We now generalize the above consideration.  Let ${\bar S}[\phi, \phi^*]$
be an action that defines a quantum system via the functional
integration over $\phi$.  Under the BRST transformation of fields
\begin{eqnarray}
\delta \phi^A = \frac{\partial^l {\bar S}}{\partial \phi^*_A},
\nn
\end{eqnarray}
the changes in the action and the functional measure are summed up to
the quantum master operator:

\begin{equation}
{\bar \Sigma}[\phi,\phi^{*}]\equiv
\frac{\partial^{r} {\bar S}}{\partial \phi^{A}}
\frac{\partial^{l}{\bar S}}{\partial \phi_{A}^*} +
\frac{\pa^r}{\pa\phi^{A}}\delta \phi^{A}
= \frac{1}{2}({\bar S},~{\bar S}) + \Delta {\bar S}\,,
\label{BV-QMoperator}
\end{equation}
where we define
\begin{equation}
\Delta \equiv (-)^{\epsilon_A +1} \frac{\pa^{r}}{\pa \phi^{A}}
\frac{\pa^{r}}{\pa \phi_{A}^{*}}\,.
\label{BV-Delta operator}
\end{equation}
The system is BRST invariant quantum mechanically if the two
contributions cancel:
\begin{equation}
\bar{\Sigma} [\phi,\phi^*] = 0\,.\label{BV-QME}
\end{equation}
We call this equation the {quantum master equation} ({QME}).

We define the quantum BRST transformation as
\begin{eqnarray}
\delta_Q X \equiv (X, {\bar S}) + \Delta X
\label{BV-qBRST}
\end{eqnarray}
for an arbitrary variable $X$.  Without assuming QME, we obtain two
important algebraic identities:
\begin{eqnarray}
&&\delta_Q {\bar \Sigma}[\phi, \phi^*]=0\,,
\label{BV-del sigma zero}\\
&&\delta_Q^2 X = (X, {\bar \Sigma}[\phi, \phi^*])\,.
\label{BV-qBRST squared}
\end{eqnarray}
These are consequences of the definitions of the quantum master operator
\bref{BV-QMoperator} and the quantum BRST transformation
\bref{BV-qBRST}.  The identity \bref{BV-del sigma zero} is crucial for
the perturbative construction of symmetric theories, as shown in \S 7
and \S 8 in Ref. \citen{review-PTPS}.  Equation~\bref{BV-qBRST squared}
implies that the quantum BRST transformation \bref{BV-qBRST} is
nilpotent if and only if QME \bref{BV-QME} holds.

\subsection{Application to ERG}

For the application of BV formalism to the exact renormalization group
(ERG), we take the action defined at some ultraviolet scale $\Lambda_0$:
\begin{eqnarray}
S_B[\phi] \equiv -\frac{1}{2}\phi \cdot K_0^{-1} D \cdot \phi +
 S_{I,B}[\phi].
\label{bare action}
\end{eqnarray}
Here, the momentum of the propagating mode is restricted as $p^2 <
\Lambda_0^2$ with a positive function $K_0(p) \equiv \kappa(p^2/\Lambda_0^2)$;
the function $\kappa$ behaves as
\begin{eqnarray}
 \quad \kappa(p^2/\Lambda^2)\sim \quad  \left\{
		\begin{array}{ll}
		 1, & (p^2/\Lambda^2 < 1) \\
		 0. & (p^2/\Lambda^2 >  1)
		\end{array}
               \right.
\label{cutoff-func}
\end{eqnarray}
We also use the following notation:
\begin{eqnarray}
\phi \cdot K_0^{-1} D \cdot \phi \equiv \int_p
 \phi^A(-p)\frac{D_{AB}(p)}{K_0(p)}\phi^B(p). \nn
\end{eqnarray}
Suppose that we have the extended action ${\bar S}_B[\phi, \phi^*]$ based
on the BRST invariance of the bare action \bref{bare action}.  Then we
define the partition function as
\begin{eqnarray}
{\bar Z}_B[J,\phi^*] \equiv \exp[{\bar W}_{B}[J,\phi^*]] \equiv \int {\cal D}\phi \exp ({\bar
 S}_B[\phi,\phi^*]+K_0^{-1}J\cdot \phi).
\label{partition function}
\end{eqnarray}
By introducing the momentum cutoff $\Lambda$, lower than the UV cutoff,
we may perform momentum integration for $\Lambda^2 < p^2 < \Lambda^2_0$.
This gives the Wilson action ${\bar
S}_{\Lambda}$ with the cutoff $\Lambda$ and the corresponding generating functional,
\begin{eqnarray}
{\bar Z}_{\Lambda}[J,\Phi^*] = \int {\cal D} \Phi \exp \Bigl(
{\bar S}_{\Lambda}[\Phi,\Phi^*]+K^{-1}J \cdot \Phi \Bigr)~,
\label{Z for Lambda}
\end{eqnarray}
where $K(p) \equiv \kappa(p^2/\Lambda^2)$.
The field $\Phi^A$ carries momentum lower than the scale $\Lambda$ and
we have rescaled the antifields as
\begin{eqnarray}
K \Phi^*_A = K_0 \phi^*_A, \label{AFs}
\label{scaling AF}
\end{eqnarray}
in order to keep the canonical structure.
The Wilson action takes the form,
\begin{eqnarray}
{\bar S}_{\Lambda}[\Phi, \Phi^*] &\equiv& - \frac{1}{2}\Phi \cdot K^{-1}D
 \cdot \Phi + {\bar S}_{I,\Lambda}[\Phi,\Phi^*]~,
\label{Wilson action}\\
\exp({\bar S}_{I,\Lambda}[\Phi,\Phi^*]) &=& \int {\cal D}\chi
\exp \Bigl[-\frac{1}{2}\chi \cdot (K_0-K)^{-1}D \cdot \chi +
{\bar S}_{I,B}[\Phi+\chi,\phi^*]
\Bigr]~.
\nn
\end{eqnarray}
Two generating functionals \bref{partition function} and \bref{Z for
Lambda} are related as
\begin{eqnarray}
{\bar Z}_B[J,\phi^*] = N_J {\bar Z}_{\Lambda}[J,\Phi^*]~,
\label{relating generating functionals}
\end{eqnarray}
where
\begin{eqnarray}
{\rm ln} N_J = - \frac{(-)^{\epsilon_A}}{2} J_A K_0^{-1}K^{-1} (K_0-
 K)(D^{-1})^{AB}J_B~.
\label{N J}
\end{eqnarray}
In this manner, we may observe the change
in the Wilson action under the change in the cutoff scale $\Lambda$.
We denote QM operators at the scales $\Lambda_0$ and $\Lambda$ as ${\bar
\Sigma}_B$ and ${\bar \Sigma}_{\Lambda}$ respectively.

Rather than following the above-mentioned standard procedure, we may take a
different way to integrate over the same momentum modes and introduce
the effective average action.  We consider the path integral
\begin{eqnarray}
\exp[{\bar W}_{B,\Lambda}[J,\phi^*]] \equiv \int {\cal D}\phi \exp
 \Bigl( {\bar S}_{B,\Lambda}[\phi,\phi^*]+K_0^{-1}J \cdot \phi \Bigr).
\label{generating functional}
\end{eqnarray}
In two path integrals, \bref{partition function} and \bref{generating
functional}, the action ${\bar S}_{B,\Lambda}$ differs from ${\bar S}_B$
only in the kinetic term: the action ${\bar S}_{B, \Lambda}$ has the kinetic term
\begin{eqnarray}
\phi \cdot (K_0-K)^{-1} D \cdot \phi~,
\label{kinetic term}
\end{eqnarray}
and two actions are related as
\begin{eqnarray}
{\bar S}_B = {\bar S}_{B,\Lambda} + \frac{1}{2} \phi \cdot R_{\Lambda}
 \cdot \phi
\label{relating two actions}
\end{eqnarray}
where
\begin{eqnarray}
[R_{\Lambda}(p)]_{BA} \equiv D_{BA}(p)\Bigl(\frac{1}{K_0-K}-\frac{1}{K_0}\Bigr)~.
\label{def R}
\end{eqnarray}
In particular, the two actions become the same in $\Lambda \rightarrow
0$.

Since the factor $K_0-K \sim 1$ for $\Lambda^2 < p^2 < \Lambda^2_0$,
while it is zero otherwise, this kinetic term allows only the modes with
$\Lambda^2 < p^2 < \Lambda^2_0$ to contribute to the path integral.
From the generating functional defined in Eq.
\bref{generating functional}, we define the effective average action as
\begin{eqnarray}
{\bar \Gamma}_{B,\Lambda}[\varphi_{\Lambda},\phi^*] \equiv {\bar W}_{B, \Lambda}[J,\phi^*]-K_0^{-1}J
 \cdot \varphi_{\Lambda}~,
\label{effective average action}
\end{eqnarray}
where
\begin{eqnarray}
K_0^{-1}\varphi_{\Lambda}(p) \equiv \frac{\partial^l {\bar W}_{B,\Lambda}[J,\phi^*]}{\partial
 J(-p)} .
\end{eqnarray}
It is the Wetterich equation, the flow of the effective average action,
that is often used for practical calculations.

In the limit of $\Lambda \rightarrow 0$, the path integral of the
r.h.s. of Eq. \bref{generating functional} reduces to that of
Eq. \bref{partition function}.  Therefore, the effective average action
is nothing but the ordinary effective action to be denoted as
${\bar \Gamma}_B[\varphi, \phi^*]$, where $\varphi \equiv \lim_{\Lambda
\rightarrow 0} \varphi_{\Lambda}$: namely,
\begin{eqnarray}
\lim_{\Lambda \rightarrow 0} {\bar
 \Gamma}_{B,\Lambda}[\varphi_{\Lambda},\phi^*] = {\bar \Gamma}_B
 [\varphi, \phi^*]~.
\end{eqnarray}

\subsection{QME and Zinn-Justin equation}

Let us introduce the path integral average of the QM operator ${\bar
\Sigma}_B[\phi,\phi^*]$
\begin{eqnarray}
\hspace{-8mm}{\bar {\Sigma}}^{1PI}_{B,\Lambda}[\varphi_{\Lambda},\phi^*]
\equiv \exp[-{\bar
 W}_{B,\Lambda}[J,\phi^*]]
\int {\cal D}\phi {\bar \Sigma}_B[\phi,\phi^*]
\exp\Bigl({\bar S}_{B,\Lambda}[\phi,\phi^*]+K_0^{-1}J \cdot \phi \Bigr).~~~
\label{defining the 1PI sigma}
\end{eqnarray}
Further rewriting the r.h.s. of Eq. \bref{defining
the 1PI sigma} in terms of the effective average action, we find
\begin{eqnarray}
{\bar {\Sigma}}^{1PI}_{B,\Lambda} [\varphi_{\Lambda}, \phi^*]&=&
\frac{\partial^r {\bar \Gamma}_{B,\Lambda}}{\partial \varphi_{\Lambda}^A}
\frac{\partial^l {\bar \Gamma}_{B,\Lambda}}{\partial \phi^*_A}
\nn\\
&{}&+[R_{\Lambda}]_{BA}
\Bigl(
-({\bar \Gamma}^{(2)})^{-1}_{B,\Lambda} \frac{\partial^l}{\partial
\varphi_{\Lambda}^C}
\frac{\partial^l {\bar \Gamma}_{B,\Lambda}}{\partial \phi^*_A}
+
\varphi_{\Lambda}^B
\frac{\partial^l {\bar \Gamma}_{B,\Lambda}}{\partial \phi^*_A}
\Bigr)~.
\label{roughly ZJ equation}
\end{eqnarray}
Since $R_{\Lambda} \rightarrow 0$ (cf. Eq. \bref{def R}) in the limit of $\Lambda \rightarrow 0$,
we find
\begin{eqnarray}
{\bar {\Sigma}}^{1PI}_{B} \equiv
\lim_{\Lambda \rightarrow 0}{\bar {\Sigma}}^{1PI}_{B,\Lambda} =
\frac{\partial^r {\bar \Gamma}_{B}}{\partial \varphi^A}
\frac{\partial^l {\bar \Gamma}_{B}}{\partial \phi^*_A}~.
\label{ZJ eq}
\end{eqnarray}

The vanishing of the quantum master operator ${\bar
       \Sigma}_B[\phi,\phi^*]=0$ implies the presence of a symmetry.
       Via \bref{defining the 1PI sigma}, this corresponds to the
       modified Slavnov-Taylor identity\cite{Ellwanger:1994iz} ${\bar
       \Sigma}^{1PI}_{B,\Lambda}=0$, which reduces to the Zinn-Justin
       equation for the effective action ${\bar \Gamma}_B$ in the limit
       of $\Lambda \rightarrow 0$.

\subsection{Flow equation and composite operator}

Under the scale change, the Wilson action changes according to the
Polchinski equation\cite{Polchinski:1983gv}
\begin{eqnarray}
&{}&-\Lambda \frac{\partial}{\partial \Lambda}{\bar S}_{\Lambda} =
\int_p (K^{-1}\Delta)(p) \Bigl[
\Phi^A(p) \frac{\partial^l {\bar S}_{\Lambda}}{\partial \Phi^A(p)}
-\Phi^*_A(p) \frac{\partial^l {\bar S}_{\Lambda}}{\partial \Phi^*_A(p)}
\Bigr]
\nn\\
&{}&~~~~+\frac{1}{2}\int_p(-)^{\epsilon_A}(D^{-1}\Delta)^{AB}(p)
\Bigl[
\frac{\partial^l {\bar S}_{\Lambda}}{\partial \Phi^B(-p)}
\frac{\partial^r {\bar S}_{\Lambda}}{\partial \Phi^A(p)}
+
\frac{\partial^l \partial^r {\bar S}_{\Lambda}}{\partial \Phi^B(-p)\partial \Phi^
A(p)}
\Bigr],
\label{Polchinski equation}
\end{eqnarray}
where
\begin{eqnarray}
\Delta(p^2/\Lambda^2) \equiv \Lambda \frac{\partial}{\partial \Lambda}
 \kappa(p^2/\Lambda^2)~.
\label{derivative of K}
\end{eqnarray}
Together with the boundary condition
\begin{eqnarray}
{\bar S}_{\Lambda} \vert_{\Lambda=\Lambda_0} = {\bar S}_B~,
\label{boundary condition}
\end{eqnarray}
the flow equation \bref{Polchinski equation} determines the Wilson
action uniquely.

We define a composite operator ${\bar {\mathcal O}}_{\Lambda}$ as a functional for
which ${\bar S}_{\Lambda}$ and its infinitesimal perturbation ${\bar
S}_{\Lambda}+ \epsilon {\bar {\mathcal O}}_{\Lambda}$ satisfy the same flow
equation \bref{Polchinski equation}.  The flow of such an operator is
given as
\begin{eqnarray}
-\Lambda \frac{\partial}{\partial \Lambda} {\bar {\mathcal O}}_{\Lambda} = {\bar
 {\mathcal D}}{\bar {\mathcal O}}_{\Lambda},
\label{flow of composite operator}
\end{eqnarray}
where
\begin{eqnarray}
{\bar {\mathcal D}} \equiv -(K^{-1}\Delta)\Phi^*_A
 \frac{\partial^l}{\partial \Phi^*_A} + (-)^{\epsilon_A}
 (D^{-1}\Delta)^{AB}
\Bigl(
\frac{\partial^l {\bar S}_{I,\Lambda}}{\partial \Phi^B}
\frac{\partial^r}{\partial \Phi^A}
+
\frac{1}{2} \frac{\partial^l \partial^r}{\partial \Phi^B \partial \Phi^A}
\Bigr) .
\label{D bar}
\end{eqnarray}

\section{QM operator and anomalies}

After these preparations, we may describe the main subject of the present
paper.  More results supporting the following arguments will be
presented in later sections.

The QM operator is a composite operator
\begin{eqnarray}
-\Lambda \frac{\partial}{\partial \Lambda} {\bar \Sigma}_{\Lambda} =
 {\bar {\mathcal D}}{\bar \Sigma}_{\Lambda}~.
\label{flow of sigma bar}
\end{eqnarray}
Thus, if the QM operator vanishes at some scale, it does so all down to
$\Lambda=0$: this is the manifestation of a symmetry.  For an anomalous
theory, however, it does not vanish and its asymptotic form in the limits
of $\Lambda \rightarrow \infty$ after taking $\Lambda_0 \rightarrow
\infty$ is an anomaly multiplied by a ghost, which will be denoted as ${\cal A}[\phi]$
\begin{eqnarray}
\lim_{\Lambda \rightarrow \infty}
\lim_{\Lambda_0 \rightarrow \infty}
{\bar \Sigma}_{\Lambda} =%
{\cal A}[\phi]~.
\label{asymptotic sigma}
\end{eqnarray}
Note that ${\cal A}[\phi]$ is written in terms of the bare field
$\phi$.  We will come back to Eq. \bref{asymptotic sigma} in a concrete example in
the next subsection.

Under the quantum BRST transformation, the QM operator vanishes at any scale
\begin{eqnarray}
\delta_Q {\bar \Sigma}_{\Lambda} =0~.
\label{delta Sigma}
\end{eqnarray}
The QM operator is a cohomologically closed operator.  This is an algebraic
relation that holds even if the QM operator does not vanish.

By Eq. \bref{defining the 1PI sigma}, we defined the 1PI counterpart for
the QM operator, which has the expression as in Eq. \bref{roughly ZJ
equation}.  In the limit of $\Lambda \rightarrow 0$, we find
Eq. \bref{ZJ eq}.  Therefore, for an anomalous theory, we would find
\begin{eqnarray}
{\bar {\Sigma}}^{1PI}_{B} \equiv
\lim_{\Lambda \rightarrow 0}{\bar {\Sigma}}^{1PI}_{B,\Lambda} =
\frac{\partial^r {\bar \Gamma}_{B}}{\partial \varphi^A}
\frac{\partial^l {\bar \Gamma}_{B}}{\partial \phi^*_A} = {\cal A}[\varphi]~.
\label{anomalous ZJ eq}
\end{eqnarray}
Here, on the r.h.s. of \bref{anomalous ZJ eq}, there appears the same
functional ${\cal A}$ as Eq. \bref{asymptotic sigma}, but
written in terms of the classical field $\varphi$.  In this limit, the
cohomological condition on ${\cal A}[\varphi]$ is the Wess-Zumino
condition,
\begin{eqnarray}
({\cal A}[\varphi], {\bar \Gamma}_B)_{\varphi,\phi^*} =0~.
\label{Wess-Zumino}
\end{eqnarray}

Let us further study the QM operator for finite $\Lambda$.  We will give
an argument for having the same functional in Eqs. \bref{asymptotic sigma} and \bref{anomalous ZJ eq}.

As explained in Appendix D of Ref. \citen{review-PTPS}, there holds
the relation between ${\bar \Sigma}_{\Lambda}$ and ${\bar
\Sigma}_{B,\Lambda}$,\footnote{Equations \bref{relating two sigmas} and
\bref{relating two fields} are different from Eqs. (D$\cdot$16) and
(D$\cdot$19) of Ref. \citen{review-PTPS} in two points: 1) here, the UV cutoff
$\Lambda_0$ is finite; 2) we rescaled the antifields according to
Eq. \bref{AFs} to respect the canonical relation.  Other than these two minor
differences, the relations are the same.}
\begin{eqnarray}
&{}&{\bar \Sigma}_{\Lambda}[\Phi,\Phi^*] = {\bar
 \Sigma}_{B,\Lambda}^{1PI}[\varphi_{\Lambda},\phi^*]~,
\label{relating two sigmas}\\
&{}&
\varphi_{\Lambda}^A
= \frac{K_0}{K} \Phi^A
+ (K_0-K)  (D^{-1})^{AB}
\frac{\partial^l \bar S_\Lambda}{\partial \Phi^B}~.
\label{relating two fields}
\end{eqnarray}
In the presence of an anomaly, we may regard the QM operator as the
composite operator, which becomes the anomaly multiplied by the
ghost in both the UV and IR limits.
Equation \bref{relating two sigmas} tells us that the operator is a functional of
$\varphi_{\Lambda}$ and $\phi^*$, where $\varphi_{\Lambda}$ is a
composite operator by itself.
Therefore, we may write the QM operator as
\begin{eqnarray}
{\bar \Sigma}_{\Lambda} = {\bar {\mathcal A}}[\varphi_{\Lambda}, \phi^*; \Lambda]~.
\label{anomaly composite operator}
\end{eqnarray}
The scale dependence of the operator originates from $\varphi_{\Lambda}$, as
well as the scale dependence of coefficients.  The latter scale dependence
is expressed by the last $\Lambda$ on the r.h.s. of
Eq. \bref{anomaly composite operator}.

We consider the flow equation for ${\bar {\mathcal A}}[\varphi_{\Lambda}, \phi^*;
\Lambda]$,
\begin{eqnarray}
-\Lambda \frac{\partial}{\partial \Lambda} {\bar {\mathcal A}} [\varphi_{\Lambda}, \phi^*;
\Lambda] = {\mathcal D}
 {\bar {\mathcal A}} [\varphi_{\Lambda}, \phi^*; \Lambda]~,
\label{flow for A}
\end{eqnarray}
where
\begin{eqnarray}
{\mathcal D} \equiv (D^{-1}\Delta)^{AB}
\Bigl( \frac{\partial^l {\bar S}_{I,\Lambda}}{\partial
\Phi^B}\frac{\partial^l}{\partial \Phi^A} + \frac{1}{2}
\frac{\partial^l}{\partial \Phi^B}\frac{\partial^l}{\partial \Phi^A}
\Bigr)~.
\label{flow operator}
\end{eqnarray}
The operator \bref{flow operator} is different from the one in
Eq. \bref{D bar} in two points: 1) in Eq. \bref{flow for A}, we do not include the trivial
scale change of the antifield given in Eq. \bref{scaling AF}; 2) the
right derivative w.r.t. $\Phi^A$ in \bref{D bar} is rewritten into the
left derivative in \bref{flow operator}.

Note that $\varphi_{\Lambda}$ is a functional of $\Phi$ and $\phi^*$ via
Eq. \bref{relating two fields} and a composite operator by itself that follows the
same flow equation \bref{flow for A} as ${\bar {\mathcal A}} [\varphi_{\Lambda}, \phi^*; \Lambda]$.
Using this fact, we may separate the scale dependence of
${\bar {\mathcal A}} [\varphi_{\Lambda}, \phi^*; \Lambda]$ into two parts from
$\varphi_{\Lambda}$ and coefficients, respectively.  The latter follows the equation
\begin{eqnarray}
\Bigl(-\Lambda \frac{\partial}{\partial \Lambda}\Bigr)'
{\bar {\mathcal A}} [\varphi_{\Lambda}, \phi^*; \Lambda]
= {\mathcal D}' {\bar {\mathcal A}} [\varphi_{\Lambda}, \phi^*; \Lambda] ,
\label{flow of C}
\end{eqnarray}
where
\begin{eqnarray}
{\mathcal D}' \equiv
 \frac{1}{2}(-)^{\epsilon_A+\epsilon_B(\epsilon_A+\epsilon_C)}
(D^{-1}\Delta)^{AB}
\Bigl(\frac{\partial^l \varphi^C_{\Lambda}}{\partial \Phi^A}
\frac{\partial^l \varphi^D_{\Lambda}}{\partial \Phi^B}\Bigr)
\frac{\partial^l}{\partial \varphi_{\Lambda}^D}
\frac{\partial^l }{\partial \varphi_{\Lambda}^C}
~.
\end{eqnarray}
The prime on the derivative on the l.h.s. of Eq. \bref{flow of C} implies that it acts on
the explicit scale dependence through coefficients.

In deriving Eq. \bref{flow of C}, we used only the relation ${\bar
{\mathcal O}}[\Phi,\Phi^*] = {\bar {\mathcal
O}}^{1PI}[\varphi_{\Lambda},\phi^*]$ for a composite operator and its
1PI counterpart.  Therefore, it is valid for any 1PI composite operator.

Now, we make the loop expansion of ${\bar {\mathcal A}}$.  Since there is
no tree-level contribution, we find
\begin{eqnarray}
\Bigl(-\Lambda \frac{\partial}{\partial \Lambda}\Bigr)'
{\bar {\mathcal A}}^{(1)} [\varphi_{\Lambda}, \phi^*; \Lambda]
= 0~
\end{eqnarray}
for the one-loop contribution.  In other words, at the one-loop level,
the scale dependence originates solely from $\varphi_{\Lambda}$~.

Let us assume for the moment that the one-loop calculation is exact.  In
the limit of $\Lambda \rightarrow 0$, the operator becomes ${\bar {\mathcal A}}
[\varphi, \phi^*]$ that must be cohomologically equivalent to the well-known
form of the anomaly denoted as ${\mathcal A}[\varphi]$ earlier in
Eq. \bref{anomalous ZJ eq}.  At this point, we realize
that there must be a composite operator that tends to ${\mathcal
A}[\varphi]$ in $\Lambda \rightarrow 0$.  It is the operator
${\mathcal A}[\varphi_{\Lambda}]$ that is the same functional as
Eq. \bref{anomalous ZJ eq} with all the fields replaced by
$\varphi_{\Lambda}$.

We may summarize our discussion that all the known facts are
consistent with the following expression for the QM operator,
\begin{eqnarray}
{\bar \Sigma}_{\Lambda}[\Phi, \Phi^*] = {\mathcal A}[\varphi_{\Lambda}]~.
\end{eqnarray}
Via the composite operator ${\mathcal A}[\varphi_{\Lambda}]$, the two limits
\bref{asymptotic sigma} and \bref{anomalous ZJ eq} are related.

\subsection{Anomaly and QM operator: $U(1)_V \times U(1)_{A}$ gauge theory}

Here, we explain how the QM operator is related to an anomaly by taking the
$U(1)_V \times U(1)_{A}$ gauge theory as an example.

Let us state a few facts that will be found useful later.

In general, the QM operator ${\bar \Sigma}_{\Lambda}[\Phi, \Phi^*]$ is related to
the Ward-Takahashi operator ${\Sigma}_{\Lambda}[\Phi]$ as
\begin{eqnarray}
{\Sigma}_{\Lambda}[\Phi] = {\bar \Sigma}_{\Lambda}[\Phi, \Phi^*]
 \vert_{\Phi^*=0} .
\nonumber
\end{eqnarray}

For QED, antifields appear in the Wilson action in a simple
manner.\cite{Igarashi:2007fw, Higashi:2007ax} As a result, the QM
operator may be obtained via shifting the fields in the WT operator:
\begin{eqnarray}
{\bar \Sigma}_{\Lambda} = \Sigma_{\Lambda}[A_{\mu}^{sh}, c, \psi^{sh},
 {\bar \psi}^{sh}],
\label{QM to WT}
\end{eqnarray}
as explained in Refs. \citen{review-PTPS} and \citen{Sonoda:2007}.
Here the superscript ``$sh$'' implies
that they are shifted by terms with antifields.  In other words, the antifield
dependence of the QM operator appears only in these shifts.  The shifted
variables are\footnote{The expressions of the shifted variables are given
in the limit of $\Lambda_0 \rightarrow \infty$.  In the rest of this
section, we assume that this limit has been taken.}
\begin{eqnarray}
A^{sh}_{\mu}(k) &\equiv& A_{\mu}(k) - \frac{1-K(k)}{k^2}k_{\mu}{\bar
 c}^*(k)~,
\nn\\
\psi^{sh}(p) &\equiv& \psi(p) + \frac{1-K(p)}{\Slash{p}+im} e
\int_k c(k) \psi^*(p-k)~,
\nn\\
{\bar \psi}^{sh}(-p) &\equiv& {\bar \psi}(-p) +
e \int_k {\bar \psi}^*(-p-k)c(k) \frac{1-K(p)}{\Slash{p}+im}~.
\label{shifted gauge}
\end{eqnarray}
For later discussion, the form of the shifted gauge field will be of
particular importance: the second term on the r.h.s. is proportional to
the momentum $k_{\mu}$.  The origin of this term is the BRST
transformation and $\phi^* \cdot \delta \phi$ term in the extended
action.

Now, let us consider a $U(1)_V \times U(1)_{A}$ gauge theory with
massless fermion with couplings
\begin{eqnarray}
\int d^4 x {\bar \psi}(e_{V} \Slash{A} + e_{A} \gamma_5 \Slash{B}) \psi .
\nn
\end{eqnarray}
For two gauge symmetries, we have WT operators, $\Sigma_{\Lambda}^V$
and $\Sigma_{\Lambda}^A$.  In Ref. \citen{review-PTPS}, their asymptotic
behaviours in $\Lambda \rightarrow \infty$ were studied.  If we keep the vector gauge symmetry intact,
$\Sigma^V_{\Lambda}=0$, we find that the WT operator for the axial
symmetry behaves as
\begin{eqnarray}
\Sigma_{\Lambda}^{A}
&\rightarrow&
- \frac{e_{A}e_{V}^{2}}{4\pi^{2}}
\epsilon_{\mu\nu\rho\sigma}
\int_{q,k}~~c_{A}(-q-k) k_{\mu}A_{\nu}(k) q_{\rho} A_{\sigma}(q)\nn\\
&& - \frac{e_{A}^{3}}{12\pi^{2}}
\epsilon_{\mu\nu\rho\sigma}
\int_{q,k}~~c_{A}(-q-k) k_{\mu}B_{\nu}(k) q_{\rho} B_{\sigma}(q)
\label{anomaly to WT operator}
\end{eqnarray}
in the limit of $\Lambda \rightarrow \infty$.  Here $c_{A}$ is the ghost
field associated with the axial gauge symmetry.

Now, we consider the QM operator, ${\bar \Sigma}_{\Lambda}^A$.  If one
recalls the reason why we find the shift in the gauge field as
Eq. \bref{shifted gauge} for QED, we understand that the same reason
applies here for both gauge fields, and ${\bar \Sigma}_{\Lambda}^A$
is written in terms of shifted gauge fields,
\begin{eqnarray}
A^{sh}_{\mu}(k) &\equiv& A_{\mu}(k) - \frac{1-K(k)}{k^2}k_{\mu}{\bar
 c}_V^*(k)~,
\nn\\
B^{sh}_{\mu}(k) &\equiv& B_{\mu}(k) - \frac{1-K(k)}{k^2}k_{\mu}{\bar
 c}_A^*(k)~.
\nn
\end{eqnarray}
The shift parts, however, vanish in $\Lambda \rightarrow \infty$.  We
conclude that the QM operator has the
same asymptotic form \bref{anomaly to WT operator} as the WT operator.

\section{Anomaly via ERG calculation}

Using the ERG approach, we explicitly calculate the anomaly contribution to
the WT operator for an abelian gauge symmetry.  We will understand where
to find anomalous contributions.  The calculations to determine counter
terms will be omitted.  It is also possible to extend the following
calculations to non-abelian gauge symmetries.\cite{Sato}

First let us sketch our calculation.  The WT operator $\Sigma_{\Lambda}$ takes the
following form
\begin{eqnarray}
\Sigma_{\Lambda} = {\bar \Sigma}_{\Lambda}\vert_{\Phi^*=0} = \frac{\partial^r
 S_{\Lambda}}{\partial \Phi^A}\delta \Phi^A+\frac{\partial^l}{\partial
 \Phi^A}\delta \Phi^A
\label{WT op}
\end{eqnarray}
where $\delta \Phi^A \equiv (\Phi^A, {\bar S}_{\Lambda})\vert_{\Phi^*=0}$~.
We will find
the second term in \bref{WT op} contains the fermion loop.  After
writing the one-loop contributions, we take $\Lambda_0 \rightarrow
\infty$ and then $\Lambda \rightarrow \infty$.  This procedure produces
an anomaly times appropriate ghost.

To evaluate the WT operator, we need to know how the BRST
transformation changes under the scale change.  For a particular class
of BRST transformation
\begin{eqnarray}
\delta \phi^A = K_0 \Bigl(
{\mathcal R}^{(1)A}_B(\Lambda_0) \phi^B + \frac{1}{2}{\mathcal R}^{(2)A}_{BC}(\Lambda_0)\phi^B\phi^C
\Bigr)
\nn
\end{eqnarray}
we have
\begin{eqnarray}
\delta \Phi^A = K
\Bigl(
{\mathcal R}^{(1)A}_B(\Lambda_0) [\Phi^B]_{\Lambda} + \frac{1}{2}{\mathcal R}^{(2)A}_{BC}(\Lambda_0)[\Phi^B\Phi^C]_{\Lambda}
\Bigr)
\label{BRST Lambda}
\end{eqnarray}
for a lower scale $\Lambda$.\cite{Igarashi:2008bb}  The
$[\Phi^B]_{\Lambda}$ and $[\Phi^B\Phi^C]_{\Lambda}$ are the composite
operators at the scale $\Lambda$:
\begin{eqnarray}
&& [\Phi^A]_{\Lambda} \equiv \Phi^A +(K_0-K)(D^{-1})^{AB}
\frac{\partial^l S_{I,\Lambda}}{\partial \Phi^B}~,
\label{comp Phi}\\
&& [ \Phi^A \Phi^B ]_{\Lambda} \equiv [\Phi^A]_{\Lambda}[\Phi^B]_{\Lambda} \nn\\
&{}&\hspace{2cm}+(K_0-K)(D^{-1})^{AC} (K_0-K)(D^{-1})^{BD}
\frac{\partial^l \partial^l S_I}{\partial \Phi^C \partial \Phi^D}~.
\label{comp Phi2}
\end{eqnarray}

Let us take again the $U(1)_V \times U(1)_{A}$ gauge theory as our example.
The interaction part of the classical action is
\begin{eqnarray}
{\cal S}_{I}[\phi]
&=& \int_{p,~k} \psibar(-p-k)\Bigl\{e_{V} \Slash{A}(k)+
e_{A} \gamma_{5}\Slash{B}(k) \Bigr\} \psi(p)
\,,
\label{cal-S-U1U1}
\end{eqnarray}
and the classical BRST transformations of fermions
and gauge fields are
\begin{eqnarray}
&&
\delta_{V} \psi(p) = -e_{V} \int_{k} \psi(p-k)c_{V}(k),~~
\delta_{V}\bar\psi(-p) = e_{V} \int_{k}\bar\psi(-p-k) c_{V}(k),~~
\nn\\
&&
\delta_{V} A_{\mu}(p) = - p_{\mu} c_{V}(p),~~
\delta_{V} B_{\mu}(p) = 0\,,
\label{vector}
\end{eqnarray}
for the vector gauge symmetry and
\begin{eqnarray}
&&
\hspace{-2mm}\delta_{A} \psi(p) = e_{A} \int_{k} \gamma_{5}\psi(p-k)c_{V}(k),~
\delta_{A}\bar\psi(-p) = e_{A} \int_{k}\bar\psi(-p-k) c_{A}(k)
\gamma_{5},
\nn\\
&&
\delta_{A} A_{\mu}(p) = 0,~~
\delta_{A} B_{\mu}(p) = - p_{\mu} c_{A}(p)\,,
\label{axial}
\end{eqnarray}
for the axial gauge symmetry.

Since transformations in Eqs. \bref{vector} and \bref{axial} are
bilinear in fields, we will have composite operators of the type
\bref{comp Phi2}.  However, the symmetries are abelian; ghosts do not
interact with other fields.  Therefore, field transformations will be
written with the fermion composite operators.  Let us take the first
transformation of \bref{vector} for example.  According to
Eq. \bref{BRST Lambda}, the transformation at the scale $\Lambda$ is
\begin{eqnarray}
\delta_V \Psi (p) = K {\cal R}^{(2)}(\Lambda_0) \int_k [\Psi]_{\Lambda}(p-k)C_V(k) ,
\label{BRST Lambda U1}
\end{eqnarray}
where
\begin{eqnarray}
[ \Psi ]_{\Lambda} (p) = \Psi(p)+\frac{K_0-K}{\Slash{p}}
\frac{\partial^l S_{I,\Lambda}}{\partial {\bar \Psi}(-p)} .
\label{comp psi}
\end{eqnarray}
To write down the interaction action $S_{I,\Lambda}$, we use
interactions in $S_{I,\Lambda_0}$ and integrate over fields with momenta
between $\Lambda_0$ and $\Lambda$.  The UV action $S_{\Lambda_0}$
contains counter terms that also affect the coefficient ${\cal
R}^{(2)}(\Lambda_0)$ in \bref{BRST Lambda U1} as well.  Here, as the
lowest order calculations, we set $\Lambda_0 = \infty$ and ignore
contributions from counter terms; we assume the classical value $-e_V$
for ${\cal R}^{(2)}(\Lambda_0)$~.\footnote{The
calculation\cite{review-PTPS} quoted in the previous subsection is based
on the asymptotic UV behaviours of the action.  Therefore, contributions
of counter terms are properly taken care of.}

Let us write down the
interaction action at the second order in couplings that are relevant
for calculating the anomaly,
\begin{eqnarray}
&&S_{I,\Lambda}[\Phi] = \nn\\
&&~~ \frac{e_{V}^{2}}{2}\int_{l,k,q}\hspace{-4mm}\bar\Psi(-l-k)
\biggl[\Slash{A}(k) \frac{(1-K)(l)}{\Slash{l}}
\Slash{A}(q)
 +  \Slash{A}(q) \frac{(1-K)(l+k-q)}{\Slash{l}+ \Slash{k}-\Slash{q}} \Slash{A}(k) \biggr]
\Psi(l-q) \nn\\
&&~~ + \frac{e_{A}^{2}}{2}\int_{l,k,q}
\hspace{-4mm}\bar\Psi(-l-k)
\biggl[\Slash{B}(k) \frac{(1-K)(l)}{\Slash{l}}
\Slash{B}(q) +  \Slash{B}(q) \frac{(1-K)(l+k-q)}{\Slash{l}+ \Slash{k}-\Slash{q}} \Slash{B}(k) \biggr]
\Psi(l-q)\nn\\
&&~~ + \frac{e_{V}e_{A}}{2}\int_{l,k,q}
\hspace{-4mm}\bar\Psi(-l-k)\biggl[\gamma_{5}\Slash{B}(k) \frac{(1-K)(l)}{\Slash{l}}
\Slash{A}(q) +
\Slash{A}(q) \frac{(1-K)(l+q-k)}{\Slash{l}+
\Slash{k}-\Slash{q}}\gamma_{5}\Slash{B}(k)\nn\\
&& \hspace{1cm}+ \Slash{A}(k)\frac{(1-K)(l)}{\Slash{l}}\gamma_{5}\Slash{B}(q)
+ \gamma_{5}\Slash{B}(q)\frac{(1-K)(l+q-k)}{\Slash{l}+
\Slash{k}-\Slash{q}}\Slash{A}(k)
\biggr]
\Psi(l-q)~.
\label{S-I1}
\end{eqnarray}

The second term of Eq. \bref{WT op} with Eqs. \bref{comp psi} and
\bref{S-I1} produces the fermion one-loop contribution to the WT
operator, to be denoted $\Sigma_{\Lambda}^{(1)}$ in the following.
\begin{eqnarray}
\Sigma_{\Lambda}^{V(1)} &=& e_{V} \int_{p,k} tr \biggl[ \frac{\pa^{l} \pa^{r}S_{I,\Lambda}}
{\pa \bar\psi(-p+k) \pa \psi(p)}U_{V}(-p,p-k)
\biggr] c_{V}(k) ,\label{Sigma V1}\\
\Sigma_{\Lambda}^{A(1)} &=& -e_{A} \int_{p,k} tr \biggl[ \frac{\pa^{l} \pa^{r}S_{I,\Lambda}}
{\pa \bar\psi(-p+k) \pa \psi(p)}U_{A}(-p,p-k)
\biggr] c_{A}(k) ,\label{Sigma A1}
\end{eqnarray}
where
\begin{eqnarray}
U_{V}(-p, p-k) &\equiv& \frac{K(p)(1-K)(p-k)}{\Slash{p}-\Slash{k}}
- \frac{K(p-k)(1-K)(p)}{\Slash{p}}~,
\label{UV1}\nn\\
U_A(-p,p-k) &\equiv& - U_V(-p,p-k)\gamma_5~.
\nn
\end{eqnarray}

Rewriting \bref{Sigma A1}, we find
\begin{eqnarray}
\Sigma^{A(1)}_{\Lambda} &=& - e_{A}e_{V}^{2} \int_{q,k} ~c_{A}(-q-k) \epsilon_{\mu\nu\rho\sigma}
k_{\mu}A_{\nu}(k) A_{\sigma}(q) I_{\rho}(q,k)\nn\\
&& - e_{A}^{3} \int_{q,k} ~c_{A}(-q-k) \epsilon_{\mu\nu\rho\sigma}
k_{\mu}B_{\nu}(k) B_{\sigma}(q) I_{\rho}(q,k)
\,,
\end{eqnarray}
where $I_{\rho}(q,k)$ stands for the integral over $P=p+q$
\begin{eqnarray}
I_{\rho}(q,k) &=& \int_{P} P_{\rho}~K(P-q) \frac{(1-K(P))(1-K(P+k))}{P^{2} (P+k)^{2}}\,,
\end{eqnarray}
which can be evaluated for $\Lambda >> q,~k$.
Expanding in the external momenta, we find in the cutoff-removed limit $\Lambda \to \infty$
\begin{eqnarray}
I_{\rho}(q,k) \rightarrow q_{\rho} \int_{P} (-2)P^{2}\frac{d K}{d P^{2}}\frac{(1-K(P))^{2}}{P^{4}}
=\frac{q_{\rho}}{24 \pi^{2}}\,.
\end{eqnarray}
Here, use has been made of the integration formula over ${\bar q}=q/\Lambda$
\begin{eqnarray}
\int_{\bar q} \frac{\Delta({\bar q}^2)(1-\kappa({\bar q}^2))^n}{{\bar q}^4} = \frac{2}{(4\pi)^2}\frac{1}{n+1}~,
\label{formula}
\end{eqnarray}
which can be proved easily.
$\Delta({\bar q}^2)$ is defined in \bref{derivative of K}.
Finally, we obtain
\begin{eqnarray}
\Sigma^{A(1)}_{\infty}&=&  - \frac{e_{A}e_{V}^{2}}{12\pi^{2}}  \epsilon_{\mu\nu\rho\sigma}
\int_{q,k}~~c_{A}(-q-k) k_{\mu}A_{\nu}(k) q_{\rho} A_{\sigma}(q)\nn\\
&&- \frac{e_{A}^{3}}{12\pi^{2}}  \epsilon_{\mu\nu\rho\sigma}
\int_{q,k}~~c_{A}(-q-k) k_{\mu}B_{\nu}(k) q_{\rho} B_{\sigma}(q)\nn\\
&=& \frac{e^{2}}{48\pi^{2}}\int_{x} c_{A}(x)
~\epsilon_{\mu\nu\rho\sigma}
 \left(F^{V}_{\mu\nu}(x)F^{V}_{\rho\sigma}(x)
+ F^{A}_{\mu\nu}(x)F^{A}_{\rho\sigma}(x)
\right)~.
\end{eqnarray}
Similarly, we find
\begin{eqnarray}
\Sigma^{V(1)}_{\infty}&=&  - 2 \times \frac{e_{A}e_{V}^{2}}{12\pi^{2}}
\epsilon_{\mu\nu\rho\sigma}
\int_{q,k}~~c_{V}(-q-k) k_{\mu}B_{\nu}(k) q_{\rho} A_{\sigma}(q)~.
\end{eqnarray}

We may add the following counter term to the Wilson action, $S_{I,\Lambda} \to S_{I,\Lambda} + S_{c}$:
\begin{eqnarray}
S_{c} = a ~\epsilon_{\mu\nu\rho\sigma}
\int_{q,k}~~B_{\mu}(k)A_{\nu}(-k-q) q_{\rho} A_{\sigma}(q) ,
\end{eqnarray}
where $a$ is a constant to be determined below. The BRST transformations
of the counter term are given as
\begin{eqnarray}
\delta_{V} S_{c} &=& -~a~
\epsilon_{\mu\nu\rho\sigma}
\int_{q,k}~~c_{V}(-q-k) k_{\mu}B_{\nu}(k) q_{\rho} A_{\sigma}(q)~,\nn\\
\delta_{A} S_{c} &=& ~a~
\epsilon_{\mu\nu\rho\sigma}
\int_{q,k}~~c_{A}(-q-k) k_{\mu}A_{\nu}(k) q_{\rho} A_{\sigma}(q)~.
\end{eqnarray}
Therefore, inclusion of the counter term $S_{I,\Lambda} \to S_{I,\Lambda} + S_{c}$
gives new contributions in $\Sigma^{A(1)}_{\infty}$ and $\Sigma^{V(1)}_{\infty}$ proportional to
$a$:
\begin{eqnarray}
\Sigma^{A(1)}_{\infty}&\to& \Sigma^{A(1)}_{\infty} = \left(a - \frac{e_{A}e_{V}^{2}}{12\pi^{2}}
 \right) \epsilon_{\mu\nu\rho\sigma}
\int_{q,k}~~c_{A}(-q-k) k_{\mu}A_{\nu}(k) q_{\rho} A_{\sigma}(q)\nn\\
&& \quad~ - \frac{e_{A^{3}}}{12\pi^{2}}\epsilon_{\mu\nu\rho\sigma}
\int_{q,k}~~c_{A}(-q-k) k_{\mu}B_{\nu}(k) q_{\rho} B_{\sigma}(q)\nn\\
\Sigma^{V(1)}_{\infty}&\to& \Sigma^{V(1)}_{\infty}=  -\left(a +  \frac{e_{A}e_{V}^{2}}{6\pi^{2}}\right)
\epsilon_{\mu\nu\rho\sigma}
\int_{q,k}~~c_{V}(-q-k) k_{\mu}B_{\nu}(k) q_{\rho} A_{\sigma}(q)
\end{eqnarray}
We choose the parameter $a$ as
\begin{eqnarray}
a = - \frac{e_{A}e_{V}^{2}}{6\pi^{2}} ,
\end{eqnarray}
so that the vector gauge symmetry is preserved.
The anomaly for the axial gauge symmetry is changed to
\begin{eqnarray}
\Sigma^{A(1)}_{\infty} &=& - \frac{e_{A}e_{V}^{2}}{4\pi^{2}}
\epsilon_{\mu\nu\rho\sigma}
\int_{q,k}~~c_{A}(-q-k) k_{\mu}A_{\nu}(k) q_{\rho} A_{\sigma}(q)\nn\\
&& - \frac{e_{A}^{3}}{12\pi^{2}}
\epsilon_{\mu\nu\rho\sigma}
\int_{q,k}~~c_{A}(-q-k) k_{\mu}B_{\nu}(k) q_{\rho} B_{\sigma}(q)~.
\end{eqnarray}
The result coincides with \bref{anomaly to WT operator}.

\section{Summary and discussion}

We have argued how the QM operator may be regarded as the anomaly
composite operator.  The operator simplifies in both ends $\Lambda
\rightarrow 0$ and $\infty$: in an intermediate scale, a composite
operator would consist of various operators if written in terms of
$\Phi$.  We have further argued that it can be written as ${\mathcal
A}[\varphi_{\Lambda}]$ for any scale if the one-loop calculation is
exact.  In the next subsection, we showed the validity of
Eq. \bref{asymptotic sigma} for an abelian theory.  We also presented a
one-loop calculation of anomaly in \S 4.

The QM operator satisfies the algebraic condition $\delta_Q {\bar
\Sigma}_{\Lambda}=0$.  Obviously, this tells us that, at any scale of
the cutoff, an anomaly is a closed form that provides us a nontrivial
element of the BRST cohomology.  Since we have the relation
\bref{relating two sigmas}, ${\bar \Sigma}_{\Lambda}^{1PI}$ also
satisfies the same condition.  However, writing it in terms of the
effective average action would not give an illuminating condition.
Observe that with a finite cutoff $\Lambda$, ${\bar
\Sigma}_{\Lambda}^{1PI}$ itself is not particularly simple.  Only in
the limit of $\Lambda \rightarrow 0$, we can show
\begin{eqnarray}
\Bigl(\frac{\partial^r {\bar \Gamma}_{B}}{\partial \varphi^A}
\frac{\partial^l {\bar \Gamma}_{B}}{\partial \phi^*_A}
,~
{\bar \Gamma}_{B}
\Bigr)_{\varphi,\phi^*}
=
e^{-{\bar W}_{B}} \int {\cal D}\phi
\Bigl(
\delta_Q {\bar \Sigma}_{B}
\Bigr)
e^{{\bar S}_{B}+K_0^{-1}J \cdot \phi}~
\label{WZ and QM}
\end{eqnarray}
after a straightforward but lengthy calculation explained in the Appendix.  The Wess-Zumino condition on the
l.h.s. is related to the condition on ${\bar \Sigma}_B$.

In earlier works,\cite{Bonini:1994xj, Bonini:1997yv, Pernici:1997ie}
anomalies have been calculated in ERG approaches.  Although the formulations
are different from ours, it was pointed out that anomalies appear in
asymptotic behaviours of operators related to the WT and QM operators in
our terminology.  The authors of Ref. \citen{Bonini:1997yv} studied
non-abelian anomaly including the evaluation of necessary counter terms.
The advantage of our formulation is the algebraic
structure of the antifield formalism.  That made our discussion more
transparent.

In the context of the renormalization group, several proofs\cite{Zee,Higashijima} were given
for the non-renormalization theorem.  Addressing the theorem in the
present framework of ERG is an important and interesting question.  We leave it for
future work.

\section*{Acknowledgements}
M. S. is grateful to M. Tanimoto for his continuous encouragement
throughout his graduate studies.  The work of K. I. is supported in part
by Grants-in-Aid for Scientific Research Nos. 21300289 and 22540270 from
the Japan Society for the Promotion of Science.

\appendix
\section{Proof of Eq. \bref{WZ and QM}}

We will show the following relation for finite cutoffs
$\Lambda$ and $\Lambda_0$:
\begin{eqnarray}
\Bigl(
{\cal A}_{B, \Lambda},~
{\bar \Gamma}_{B,\Lambda}
\Bigr)_{\varphi_{\Lambda},\phi^*}
=
e^{-{\bar W}_{B, \Lambda}} \int {\cal D}\phi
\Bigl(
\delta'_Q {\bar \Sigma}_{B,\Lambda}
\Bigr)
e^{{\bar S}_{B,\Lambda}+K_0^{-1}J \cdot \phi}~.
\label{Wess-Zumino and QM}
\end{eqnarray}
Let us explain the notations.  ${\mathcal A}_{B,\Lambda}$ stands for
the quantity
\begin{eqnarray}
{\mathcal A}_{B,\Lambda} \equiv
\frac{\partial^r {\bar \Gamma}_{B,\Lambda}}{\partial \varphi_{\Lambda}^A}
\frac{\partial^l {\bar \Gamma}_{B,\Lambda}}{\partial \phi^*_A}
=
e^{-{\bar W}_{B, \Lambda}} \int {\cal D}\phi
~{\bar \Sigma}_{B,\Lambda}
e^{{\bar S}_{B,\Lambda}+K_0^{-1}J \cdot \phi}~,
\label{Defining A}
\end{eqnarray}
where the second equality was shown in Ref. \citen{review-PTPS}.
$\delta'_Q$ and ${\bar \Sigma}_{B,\Lambda}$ are the BRST
transformation and QM operator defined with the action ${\bar
S}_{B,\Lambda}$ respectively:
\begin{eqnarray}
\delta'_Q X &\equiv& (X, {\bar S}_{B, \Lambda})+\Delta X~,
\label{delta prime}\\
 {\bar \Sigma}_{B, \Lambda} &\equiv& \frac{1}{2}({\bar S}_{B, \Lambda},
  {\bar S}_{B, \Lambda}) + \Delta {\bar S}_{B, \Lambda}~.
\label{Sigma B L}
\end{eqnarray}
It is clear from Eq. \bref{relating two actions} that the difference between
${\bar S}_B$ and ${\bar S}_{B,\Lambda}$ vanishes in $\Lambda \rightarrow
0$.  Therefore, in this limit,
\begin{eqnarray}
 \delta_Q' \rightarrow \delta_Q,~~~{\bar \Sigma}_{B,\Lambda} \rightarrow
  {\bar \Sigma}_B~.
\nn
\end{eqnarray}
By definitions given in Eqs. \bref{generating functional} and \bref{effective average action}, we also have
\begin{eqnarray}
{\bar W}_{B,\Lambda} \rightarrow {\bar W}_B,~~~{\bar \Gamma}_{B,\Lambda}
 \rightarrow {\bar \Gamma}_B
\end{eqnarray}
in the same limit.  Sending $\Lambda \rightarrow 0$ in
Eq. \bref{Wess-Zumino and QM}, we find Eq. \bref{WZ and QM}.

Now, we give a proof of \bref{Wess-Zumino and QM}.
The bracket on the l.h.s. is defined with respect to $\varphi_{\Lambda}$ and
       $\phi^*$,
\begin{eqnarray}
({\cal A}_{B, \Lambda}, {\bar \Gamma}_{B, \Lambda})_{\varphi_{\Lambda}, \phi^*}
=
\frac{\partial^r {\cal A}_{B, \Lambda}}{\partial \varphi_{\Lambda}^B}\bigg|_{\phi^*}
\frac{\partial^l {\bar \Gamma}_{B, \Lambda}}{\partial \phi^*_B}
-
\frac{\partial^r {\cal A}_{B, \Lambda}}{\partial \phi^*_B}\bigg|_{\varphi_{\Lambda}}
\frac{\partial^l {\bar \Gamma}_{B, \Lambda}}{\partial \varphi_{\Lambda}^B}~.
\end{eqnarray}

Using the last expression of Eq. \bref{Defining A}, we may regard ${\cal
A}_{B,\Lambda}$ as a functional of $J$ and $\phi^*$, ${\cal A}_{B, \Lambda}={\cal
A}_{B, \Lambda}(J, \phi^*)$.  Since the source $J^A$ is a functional of
$\varphi_{\Lambda}$ and $\phi^*$ via the relation
\begin{eqnarray}
J^A = - K_0 \frac{\partial^r {\bar \Gamma}_{B, \Lambda}}{\partial \varphi_{\Lambda}^A}~,
\end{eqnarray}
${\cal A}_{B, \Lambda}$ depends on $\varphi_{\Lambda}$ and $\phi^*$ as
 \begin{eqnarray}
{\cal A}_{B, \Lambda} = {\cal A}_{B, \Lambda}(J(\varphi_{\Lambda},
 \phi^*), \phi^*) .
\end{eqnarray}
Therefore, we find
\begin{eqnarray}
({\cal A}_{B, \Lambda}, {\bar \Gamma}_{B, \Lambda})_{\varphi_{\Lambda}, \phi^*}
&=&
\Bigl({\cal A}_{B, \Lambda}(J(\varphi_{\Lambda}, \phi^*), \phi^*), {\bar \Gamma}_{B, \Lambda}\Bigr)_{\varphi_{\Lambda},
\phi^*}
\nn\\
&=&
\frac{\pa^r {\cal A}_{B, \Lambda}}{\pa J^A}(J^A, {\bar \Gamma}_{B, \Lambda})_{\varphi_{\Lambda},
\phi^*}
-
\frac{\pa^r {\cal A}_{B, \Lambda}}{\pa \phi^*_B}\bigg|_J
\frac{\partial^l {\bar \Gamma}_{B, \Lambda}}{\partial \varphi_{\Lambda}^B}~\nn\\
&=&
-
\frac{\pa^r {\cal A}_{B, \Lambda}}{\pa \phi^*_B}\bigg|_J
\frac{\partial^l {\bar \Gamma}_{B, \Lambda}}{\partial \varphi_{\Lambda}^B}~,
\label{Expression 1}
\end{eqnarray}
since the first term of the second line vanishes
\begin{eqnarray}
\frac{\pa^r {\cal A}_{B, \Lambda}}{\pa J^A}(J^A, {\bar \Gamma}_{B,
 \Lambda})_{\varphi_{\Lambda},\phi^*}
= K_0^2 (-)^{\epsilon_A \epsilon_B}
\frac{\pa^r {\cal A}_{B, \Lambda}}{\pa J^A}\frac{\pa^r {\cal A}_{B, \Lambda}}{\pa J^B} ({\bar
\Gamma}^{(2)}_{B,\Lambda})_{AB}=0~,
\label{rel1}
\end{eqnarray}
where
\begin{eqnarray}
({\bar \Gamma_{B,\Lambda}}^{(2)})_{AB} \equiv \frac{\partial^l}{\partial
 \varphi_{\Lambda}^A}\frac{\partial {\bar \Gamma}_{B,\Lambda}}{\partial \varphi_{\Lambda}^B}~.
\nn
\end{eqnarray}
Equation \bref{rel1} is easily understood once we notice the following symmetric properties,
\begin{eqnarray}
\epsilon \Bigl({\bar \Gamma}_{AC}^{(2)}\Bigr) = \epsilon_A +
 \epsilon_C,~~~~~
%
{\bar \Gamma}^{(2)}_{CA} = (-)^{\epsilon_A+\epsilon_C+\epsilon_A \epsilon_C}
{\bar \Gamma}^{(2)}_{AC}~.
\nn
\end{eqnarray}

In calculating the $\phi^*$-derivative of ${\cal A}_{B, \Lambda}$ on the r.h.s. of
       Eq. \bref{Expression 1}, we use the path integral expression
       in Eq. \bref{Defining A}.  The derivative acting on the
       factor $\exp(-{\bar W}_{B,\Lambda})$ produces the term
\begin{eqnarray}
\frac{\pa^l {\bar W}_{B,\Lambda}}{\pa \phi^*_B}{\cal A}_{B, \Lambda} \cdot
\frac{\partial^l {\bar \Gamma}_{B, \Lambda}}{\partial \varphi_{\Lambda}^B}
&=&
{\cal A}_{B, \Lambda}(-)^{\epsilon_A+1}
\frac{\pa^l {\bar \Gamma}_{B, \Lambda}}{\pa \varphi_{\Lambda}^B}
\frac{\pa^l {\bar W}_{B, \Lambda}}{\pa \phi^*_B}
\nn\\
&=& - {\cal A}_{B, \Lambda}
\frac{\pa^r {\bar \Gamma}_{B, \Lambda}}{\pa \varphi_{\Lambda}^B}
\frac{\pa^l {\bar W}_{B, \Lambda}}{\pa \phi^*_B}
=
-{\cal A}_{B, \Lambda}^2
=0~.
\end{eqnarray}
Here, use was made of the relation in the second line,
\begin{eqnarray}
\frac{\partial^l {\bar \Gamma}_{B,\Lambda}}{\partial \phi^*_A} =
\frac{\partial^l {\bar W}_{B,\Lambda}}{\partial \phi^*_A}+
\frac{\partial^l {\bar J}_C}{\partial \phi^*_A}
\Bigl(
\frac{\partial^l {\bar W}_{B,\Lambda}}{\partial J^C} - K_0^{-1}\varphi_{\Lambda}^C
\Bigr)
=\frac{\partial^l {\bar W}_{B,\Lambda}}{\partial \phi^*_A}~.
\nn
\end{eqnarray}
Thus, we may only consider the $\phi^*$-derivative of the expression under the
       path integral in Eq. \bref{Defining A}.
\begin{eqnarray}
\hspace{-7mm}({\cal A}_{B, \Lambda}, {\bar \Gamma}_{B, \Lambda})_{\varphi_{\Lambda}, \phi^*}
&=&
- e^{-{\bar W}_{B, \Lambda}}
\frac{\pa^r}{\pa \phi^*_B}\Bigl(
\int {\cal D}\phi  {\bar \Sigma}_{B,\Lambda}
e^{{\bar S}_{B, \Lambda}+K_0^{-1}J\cdot \phi}
\Bigr)\frac{\pa^r {\bar \Gamma}_{B,\Lambda}}{\pa \varphi_{\Lambda}^B}(-)^{\epsilon_B}
\nn\\
&=&
 e^{-{\bar W}_{B, \Lambda}}
\frac{\pa^r}{\pa \phi^*_B}\Bigl(
\int {\cal D}\phi  {\bar \Sigma}_{B,\Lambda}
e^{{\bar S}_{B, \Lambda}}
\frac{\pa^l}{\pa \phi^B}e^{K_0^{-1}J\cdot \phi}
\Bigr)
\nn\\
&=&
-(-)^{\epsilon_B}
e^{-{\bar W}_{B, \Lambda}}
\int {\cal D}\phi
\Bigl[\frac{\pa^r}{\pa \phi^*_B}\frac{\pa^l}{\pa \phi^B}\Bigl({\bar \Sigma}_{B,\Lambda}e^{{\bar S}_{B, \Lambda}}\Bigr)\Bigr]
e^{K_0^{-1}J\cdot \phi}.
\label{cal1}
\end{eqnarray}
In the second line of \bref{cal1}, we rewrote the factor $\frac{\pa^l {\bar
       \Gamma}_{B,\Lambda}}{\pa \varphi^B}$ as the source $J^B$, then
       the field derivative under the path integral, and finally performed
       the partial integration.  The quantity on the third line of \bref{cal1}
       may be rewritten as
\begin{eqnarray}
-(-)^{\epsilon_B}
\Bigl[\frac{\pa^r}{\pa \phi^*_B}\frac{\pa^l}{\pa \phi^B}\Bigl({\bar
\Sigma}_{B,\Lambda}e^{{\bar S}_{B, \Lambda}}\Bigr)\Bigr]= e^{{\bar
S}_{B, \Lambda}}
(\delta_Q {\bar \Sigma}_{B,\Lambda}+{\bar \Sigma}_{B,\Lambda}^2) .
\label{cal2}
\end{eqnarray}
Here, the second term on the r.h.s. vanishes since ${\bar \Sigma}_{B,
      \Lambda}$ is Grassmann odd.  Substituting Eq. \bref{cal2} to the
      last expression of Eq. \bref{cal1}, we reach the announced result
      \bref{Wess-Zumino and QM}.



\begin{thebibliography}{99}

\bibitem{Arnone:2006ie}
S.~Arnone, T.~R.~Morris and O.~J.~Rosten,
\JL{Fields Inst. Commun.,50,2007,1}.

\bibitem{Becchi:1996an}
C.~Becchi,
hep-th/9607188.

\bibitem{Igarashi:1999rm}
Y.~Igarashi, K.~Itoh and H.~So,
\PLB{479,2000,336}.

\bibitem{Batalin:1981jr}
I.~A. Batalin and G.~A. Vilkovisky,
\PLB{102,1981,27}.

\bibitem{review-PTPS}
Y.~Igarashi, K.~Itoh and H.~Sonoda,
\PTPS{181,2010,1}.

\bibitem{Amorim:1996xp}
R.~Amorim and N.~R.~F.~Braga,
\PRD{57,1998,1225}.

\bibitem{Amorim:1998mm}
R.~Amorim, N.~R.~F.~Braga and M.~Henneaux,
\PLB{436,1998,125}.

\bibitem{Ellwanger:1994iz}
U.~Ellwanger,
\PLB{335,1994,364}.

\bibitem{Polchinski:1983gv}
J.~Polchinski,
\NPB{231,1984,269}.

\bibitem{Igarashi:2007fw}
Y.~Igarashi, K.~Itoh and H.~Sonoda,
\PTP{118,2007,121}.

\bibitem{Higashi:2007ax}
T.~Higashi, E.~Itou and T.~Kugo,
\PTP{118,2007,1115}.

\bibitem{Sonoda:2007}
H.~Sonoda,
arXiv:0710.1662.

\bibitem{Sato}
K.~Itoh and M.~Sato, unpublished.

\bibitem{Igarashi:2008bb}
Y.~Igarashi, K.~Itoh and H.~Sonoda,
\PTP{120,2008,1017}.

\bibitem{Bonini:1994xj}
M.~Bonini, M.~D'Attanasio and G.~Marchesini,
\PLB{329,1994,249}.

\bibitem{Bonini:1997yv}
M.~Bonini and F.~Vian,
\NPB{511,1998,479}.

\bibitem{Pernici:1997ie}
M.~Pernici, M.~Raciti and F.~Riva,
\NPB{520,1998,469}.

\bibitem{Zee}
A.~Zee,
\PRL{29,1972,1198}.

\bibitem{Higashijima}
K.~Higashijima, K.~Nishijima and M.~Okawa,
\PTP{67,1982,668}.

\end{thebibliography}
\end{document}